\documentclass[twocolumn,showpacs,aps,pra]{revtex4}
\usepackage{epsfig}

\begin{document}

\title{Effect of cold collisions on spin coherence and resonance shifts in a magnetically trapped ultra-cold gas}
\author{D.~M. Harber, H.~J. Lewandowski, J.~M. McGuirk\cite{qpdNIST}, and E.~A. Cornell\cite{qpdNIST}}
\affiliation{JILA, National Institute of Standards and Technology and Department of Physics, \\
University of Colorado, Boulder, Colorado 80309-0440}
\date{\today}

\begin{abstract}
We have performed precision microwave spectroscopy on ultra-cold $^{87}$Rb confined in a magnetic trap, both above and below the Bose-condensation transition.  The cold collision frequency shifts for both normal and condensed clouds were measured, which allowed the intra- and inter-state density correlations (characterized by sometimes controversial ``factors of 2") to be determined.  Additionally, temporal coherence of the normal cloud was studied, and the importance of mean-field and velocity-changing collisions in preserving coherence is discussed.
\end{abstract}

\maketitle

With the advent of modern cooling and trapping techniques, the cold collision regime has become readily accessible.  In the cold collision regime quantum statistical effects due to particle indistinguishability dominate scattering processes.  The symmetrization requirement for identical particles in an ultra-cold Bose gas enhances the probability of finding two particles nearby, causing density fluctuations.  At lower temperatures the statistical nature of the Bose gas causes the phenomena of Bose-Einstein condensation, where all atoms in the condensate share the same wavefunction, suppressing the density fluctuations found in a noncondensed sample.

Suppression of second-order density fluctuations in a condensate has been measured through analysis of the expansion energy of condensates \cite{holland1997,ketterle1997}.  In a separate experiment, the suppression of third-order density fluctuations was probed by comparing the three-body loss rate of a condensate to that of a normal cloud \cite{jila1997}.  The effect of cold collisions has also been measured as a density-dependent energy shift in atomic fountain clocks \cite{Gibble1993,Fertig2000,Sortais2000}.  These shifts are quite small ($\sim$0.1-10 mHz) due to the low densities at which the clocks operate, but are measurable because of their high precision.  The uncertainty associated with these collisional shifts can be problematic; in fact the next generation of atomic fountain clocks are based on $^{87}$Rb rather than $^{133}$Cs because the collisional shift of $^{87}$Rb is $\sim30$ times smaller.  In recent ultra-cold hydrogen experiments the cold collision shift provided the signature of Bose-condensation; below the transition a large frequency shift of the 1S-2S transition was seen, reflecting the high density of the condensate \cite{Hcoldc}.

In this paper we report precision microwave spectroscopy performed on ultra-cold and condensed $^{87}$Rb atoms confined in a magnetic trap.  Due to the high densities achievable in a magnetic trap, the collisional energy shifts were $10^5$ greater than those in $^{87}$Rb atomic clocks, allowing a high-precision measurement of the shifts of the magnetically trappable states to be made with relative ease.  The collisional shifts for both a normal and condensed sample were measured, providing a useful probe of the quantum statistics of the system.  Additionally, magnetic confinement permits long interrogation times, allowing us to characterize temporal coherence of the normal cloud under various experimental conditions.  Comparison of measured coherence times with a collisionless numerical simulation suggests that collisions preserve coherence in normal clouds.

Precision spectroscopy of trapped samples is difficult because atom trapping relies on spatial inhomogeneity of the atomic energy levels.  Spatial inhomogeneity of the energy levels broadens the transition frequency, thus limiting the precision attainable through spectroscopy.  In other work, this difficulty has been avoided by confining atoms in a blue-detuned optical dipole trap, where atoms spend little time interacting with the trapping fields \cite{chu1995}.  In this work spatial inhomogeneity of the transition frequency was minimized through the use of a pair of energy levels which experience the same trapping potential.  At a magnetic field of $\sim3.23$ G, the $|1\rangle\equiv|F=1,m_{f}=-1\rangle$ and $|2\rangle\equiv|F=2,m_{f}=1\rangle$ hyperfine levels of the $5\mbox{S}_{1/2}$ ground state of $^{87}$Rb experience the same first-order Zeeman shift.  For a normal cloud at 500 nK, each energy level is Zeeman shifted by $\sim$10 kHz across the extent of the cloud; however, at 3.23 G the \textit{differential} shift of the two levels across the cloud is $\sim$1 Hz.  Compared to the differential Zeeman shift, the energy shift due to cold collisions is then a relatively large effect at high densities, making measurements of collisional shifts in this system possible.  The small inhomogeneity allows for long coherence times, $\sim$2 seconds and longer for low-density clouds, making this system attractive for precision measurements as well as for the study of condensate coherence in the presence of a thermal cloud.

The experimental setup has been previously described \cite{lewan2002} and will be briefly summarized here.  Approximately $10^{9}$ $^{87}$Rb atoms are loaded into a vapor cell magneto-optical trap (MOT).  The atoms are then optically pumped into the $|F=1\rangle$ state by turning off the repump beam while MOT beams remain on.  Then the trapping beams are turned off and the MOT coils are ramped to a high current forming a 250 G/cm gradient to trap $|1,-1\rangle$ atoms in the quadrupole field of the coils.  The quadrupole coils are mounted on a linear servo-motor controlled track which then moves the coils 44 cm, from the MOT region to a Ioffe-Pritchard trap in the ultra-high vacuum region of the system.  The Ioffe-Pritchard trap consists of two permanent magnets which provide a 450 G/cm radial gradient.  Two pairs of electromagnetic coils, a pinch and a bias, provide confinement in the axial direction, which is aligned perpendicular with respect to gravity.  At a typical bias field of 3.23 G, atoms in the $|1,-1\rangle$ state experience $\{230,230,7\}$ Hz trap frequencies.  The sample is further cooled by rf evaporation, and condensates of up to $10^{6}$ atoms can be formed.  Imaging is performed by the use of adiabatic rapid passage to transfer atoms from the $|1,-1\rangle$ state to the $|2,-2\rangle$ state.  Anti-trapped $|2,-2\rangle$ atoms rapidly expand for 2-5 ms and then are imaged through absorption by a 20 $\mu\mbox{s}$ pulse of $5\mbox{S}_{1/2}$ $|2,-2\rangle \rightarrow 5\mbox{P}_{3/2}$ $|3,-3\rangle$ light.

A two-photon microwave-rf transition is used to transfer atoms between the $|1\rangle$ and $|2\rangle$ states.  A  detuning of 0.7 MHz from the $|2,0\rangle$ intermediate state provides a two-photon Rabi frequency of $\sim2.5$ kHz.  Ramsey spectroscopy of the $|1\rangle \rightarrow |2\rangle$ transition is performed by measuring the total number of atoms remaining in state $|1\rangle$ after a pair of $\frac{\pi}{2}$ pulses separated by a variable time delay are applied \cite{ramsey1956}.  The frequency of the resulting Ramsey fringes is the difference between the transition frequency $\nu_{12}$ and the two-photon drive frequency.  In previous work we measured local variations of $\nu_{12}$  by detecting the number of atoms remaining in state $|1\rangle$ at specific spatial locations along the axis of the normal cloud \cite{lewan2002}.  By analyzing the spatio-temporal variations of $\nu_{12}$, combined with the measured evolution of the $|1\rangle$ state after a single $\frac{\pi}{2}$ pulse, we were able to spatially resolve the evolution of spin waves \cite{mcguirk2002}.  In this work, in order to perform measurements of $\nu_{12}$ insensitive to spin waves, one of the following two techniques was used.  With one technique the entire cloud, rather than specific spatial locations, was monitored to average out the effects of spin waves.  Alternatively, Ramsey spectroscopy was restricted to interrogation times short compared to the spin wave frequency \cite{fspinwave}.

\begin{figure}
\leavevmode
\epsfxsize=3.375in
\epsffile{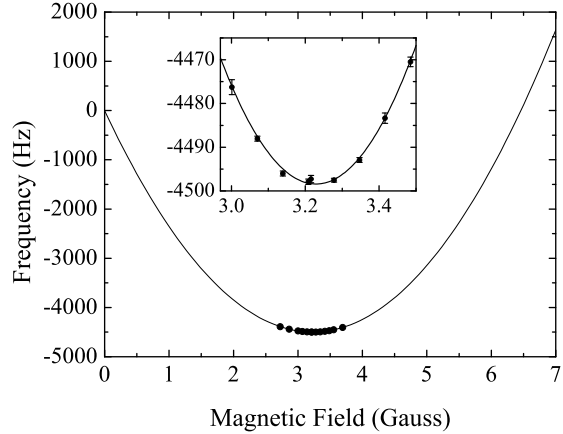} 
\caption{\label{fig:zeeman}  Differential Zeeman shift at low magnetic fields for the $|1\rangle \rightarrow |2\rangle$ ($|F=1,m_{f}=-1\rangle \rightarrow |F=2,m_{f}=1\rangle$) transition.  The solid line is the predicted splitting from the Breit-Rabi formula.  The inset plot expands the bias field region where most studies are performed.}
\end{figure}

One effect which shifts the transition frequency $\nu_{12}$ is the differential Zeeman shift.  The Breit-Rabi formula predicts a minimum in $\nu_{12}$ at $B_{0}=3.228 917(3)$ Gauss, thus the differential Zeeman shift between the $|1\rangle$ and $|2\rangle$ energy levels is first-order magnetic field independent at $B=B_{0}$.  The differential Zeeman shift about $B_{0}$ can be approximated as $\nu_{12} = \nu_{min}+\beta (B-B_{0})^{2}$ \cite{brettrabi}.  Measuring $\nu_{12}$ for different magnetic fields allows us to calibrate our magnetic field from the expected dependence, see Fig.~\ref{fig:zeeman}.  By working in the vicinity of $B=B_{0}$ we greatly reduce spatial inhomogeneity of $\nu_{12}$ and also become first-order insensitive to temporal magnetic field fluctuations.

A second effect which shifts $\nu_{12}$ arises from atom-atom interactions.  In the s-wave regime, where the thermal de Broglie wavelength of the atoms is greater than their scattering length, atoms experience an energy shift equal to $\alpha\frac{4\pi\hbar^{2}}{m}an$, where $\alpha$ is the two-particle correlation at zero separation, $n$ is atom number density, $a$ is the scattering length, and $m$ is the atom mass.  Therefore for a two-component sample the expected energy shift of each state is

\begin{eqnarray}
\delta \mu_{1} = \frac{4\pi\hbar^{2}}{m}(\alpha_{11}a_{11}n_{1} + \alpha_{12}a_{12}n_{2})\\
\delta \mu_{2} = \frac{4\pi\hbar^{2}}{m}(\alpha_{12}a_{12}n_{1} + \alpha_{22}a_{22}n_{2}),
\end{eqnarray}

\noindent where $n_1$ and $n_2$ are the $|1\rangle$ and $|2\rangle$ state density and

\begin{equation}
\alpha_{ij} = \frac{\langle\Psi^{\dag}_{i}\Psi^{\dag}_{j}\Psi_{i}\Psi_{j}\rangle}{\langle\Psi^{\dag}_{i}\Psi_{i}\rangle\langle\Psi^{\dag}_{j}\Psi_{j}\rangle}\mbox{.}
\end{equation}

\noindent The shift of the transition frequency in Hz can then be written as

\begin{eqnarray}
\Delta \nu_{12} & = & \frac{2\hbar}{m}(\alpha_{12}a_{12}n_{1} + \alpha_{22}a_{22}n_{2}-\alpha_{11}a_{11}n_{1} - \alpha_{12}a_{12}n_{2}) \nonumber\\
 & = & \frac{\hbar}{m}n(\alpha_{22}a_{22}-\alpha_{11}a_{11} + \nonumber\\
 &   & (2\alpha_{12}a_{12}-\alpha_{11}a_{11}-\alpha_{22}a_{22})f),
\end{eqnarray}

\noindent where $f=\frac{n_1-n_2}{n}$ and $n=n_1+n_2$.

For noncondensed, indistinguishable bosons, $\alpha=2$ due to exchange symmetry, therefore $\alpha_{11}^{nc}=\alpha_{22}^{nc}=2$ in a cold normal cloud (where the superscript $c$ or $nc$ refers to condensed of noncondensed atoms respectively).  Distinguishable particles do not maintain exchange symmetry, making $\alpha_{12}^{nc}=1$ for an incoherent two-component mixture.  However if a two-component sample is prepared by coherently transferring atoms from a single component, such as in Ramsey spectroscopy, then the excitation process maintains exchange symmetry, and we might expect $\alpha_{12}^{nc}=2$ \cite{Kleppner2002}.  In this scenario the collisional shift should be calculated using $\alpha_{11}^{nc}=\alpha_{22}^{nc}=\alpha_{12}^{nc}=2$, leading to a predicted frequency shift of

\begin{equation}
\Delta \nu_{12} = \frac{2\hbar}{m}n(a_{22}-a_{11} + (2a_{12}-a_{11}-a_{22})f).
\end{equation}

\noindent This result can also be obtained by solving the transport equation \cite{levitov2002,williams2002}.   From spectroscopic studies \cite{verhaar2002} the three $^{87}$Rb scattering lengths of interest have been determined to be $a_{22}=95.47a_0$, $a_{12}=98.09a_0$, and $a_{11}=100.44a_0$, where $a_0$ is the Bohr radius.  The frequency shift can then be written as

\begin{equation}
\Delta \nu_{12}= \frac{2\hbar}{m}a_{0}n(-4.97+0.27f).
\end{equation}

\noindent If on the other hand the $|1\rangle$ and $|2\rangle$ states do \textit{not} maintain exchange symmetry, such that $\alpha_{12}^{nc}=1$, then the frequency shift would instead be

\begin{equation}
\Delta \nu_{12}= \frac{2\hbar }{m}a_{0}n(-4.97-97.82f).
\end{equation}

\noindent These two models are clearly distinguished by the dependence of $\nu_{12}$ on $f$.

When we perform Ramsey spectroscopy with a pair of $\frac{\pi}{2}$ pulses, the populations of the $|1\rangle$ and $|2\rangle$ states are equal, and thus $f=0$ during the interrogation time.  From Eq.~(4) it is apparent that with $f=0$ the collisional shift is sensitive only to $\alpha^{nc}_{ii}$ and $a_{ii}$ terms.  For these measurements the bias field was set to $B_0$, and the transition frequency was measured for a range of densities.  To adjust density of the sample, the number of atoms in the initial MOT load was varied.  All normal cloud data was taken at the same temperature of 480 nK, and all condensate data was taken with high condensate fractions in order to minimize effects due to the normal cloud.  The density for the normal cloud was found by fitting Gaussian profiles to absorption images of the clouds and extracting the number, temperature, and density.  To measure condensate density, Thomas-Fermi profiles were fit to absorption images of the condensates and the total number, $N_0$, in the condensates and the Thomas-Fermi radius along the long axis, $Z$, were extracted.

\begin{figure}
\leavevmode
\epsfxsize=3.375in
\epsffile{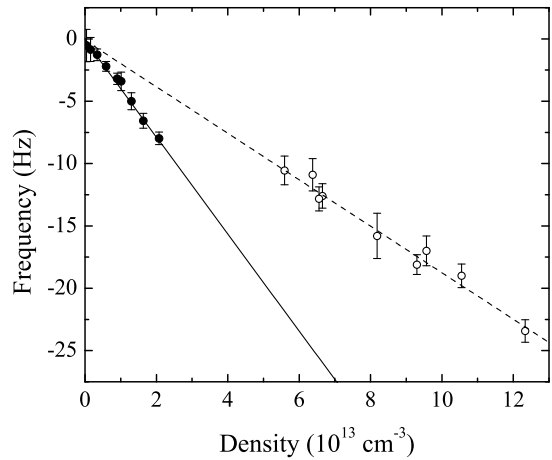} 
\caption{\label{fig:dshift}  Measurement of the cold collision shift.  Solid and open circles represent measurements of the normal cloud and condensate respectively.  The solid line is a fit to the normal cloud data $\Delta \nu_{12} = 0.1(0.4) - 3.9(0.3) 10^{-13} n$; the dashed line is a fit to the condensate data $\Delta \nu_{12} = -0.1(1.4) - 1.9(0.2) 10^{-13} n$ where $\Delta \nu_{12}$ is in Hz and $n$ is in cm$^{-3}$.}
\end{figure}

The results of this measurement are shown in Fig.~\ref{fig:dshift}.  Comparing the collisional shift measured for the normal cloud to that measured for a condensate gives $\alpha^{nc}_{ii}/\alpha^{c}_{ii}=2.1(2)$.  If instead we \textit{assume} $\alpha^{nc}_{ii}=2$ and $\alpha^{c}_{ii}=1$, then the data for both the condensate and normal cloud can be used to obtain a value for the difference in scattering lengths of $a_{22}-a_{11}=-4.92(28)a_0$, in agreement with values determined from molecular spectroscopy \cite{verhaar2002}.

Many systematics can plague density measurements made through absorption imaging.  In order to test independently our density calibration for both the normal and condensed samples, each of which can suffer from different errors, we used the Bose-condensation phenomenon.  The density of normal clouds was tested through measurement of the critical temperature, and condensate density was tested with the Thomas-Fermi approximation.  Assuming that disagreements are due only to errors in estimation of atom number, the worst case scenario, leads us to reduce normal cloud density by $11(4)\%$ and increase condensate density by $11(3)\%$.  Adjusting the cold collision shifts accordingly would yield a worst-case corrected value of $\alpha^{nc}_{ii}/\alpha^{c}_{ii}=1.7(2)$ \cite{calibration}.  The adjusted normal and condensate density shifts can be combined as above to give a value for the difference in scattering lengths of $a_{22}-a_{11}=-4.85(31)a_0$; not significantly different from our unadjusted measurement.

The remaining significant systematic is atom loss due to $|2\rangle$-$|2\rangle$ collisional dipolar relaxation.  In order to minimize effects of this loss, interrogation times were kept as short as possible.  Nevertheless for the highest density condensate measurements the $|2\rangle$-$|2\rangle$ loss causes the total density to drop by 3\% in 20 ms, the maximum interrogation time.  For all other densities the loss was no larger than this, and in most cases much smaller.  Finally, the $|1\rangle$ and $|2\rangle$ states begin undergoing spatial separation in the condensate after the first $\frac{\pi}{2}$ pulse \cite{hall1998}; however, the timescale for the separation is much longer than our 20 ms interrogation time.

\begin{figure}
\leavevmode
\epsfxsize=3.375in
\epsffile{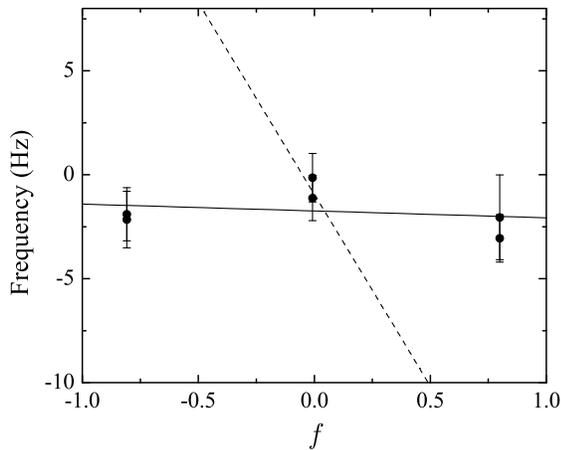} 
\caption{\label{fig:fo2}  Testing the exchange symmetry between the $|1\rangle$ and $|2\rangle$ states.  The transition frequency is measured as $f=\frac{n_1-n_2}{n}$ is varied for a normal cloud at fixed peak density of $7\times10^{12}$ cm$^{-3}$ and temperature of 510 nK.  The solid line is the fit, which yields $\alpha^{nc}_{12}/\alpha^{nc}_{11,22}=1.01(2)$, which is to say, inter- and intra-state density correlations are quite accurately the same.  The dotted line indicates the expected slope for $\alpha^{nc}_{12}/\alpha^{nc}_{11,22}=1/2$.}
\end{figure}

Exchange symmetry between the $|1\rangle$ and $|2\rangle$ states can be tested by working at a fixed density and varying the relative $|1\rangle$ to $|2\rangle$ population by varying the length of the first Ramsey pulse \cite{ExchangePulse}.  In this case the first term in Eq.~(4) will be constant and the measurement will test $\alpha^{nc}_{12}$ and $a_{12}$ as well as the $\alpha^{nc}_{ii}$ and $a_{ii}$ terms (see Eq.~(6) and (7)).  To minimize systematics the interrogation times were kept short, making precise frequency determination difficult.  Nevertheless, our measurement (Fig.~\ref{fig:fo2}) indicates $\alpha^{nc}_{12}/\alpha^{nc}_{11,22}=1.01(2)$, where we have used the spectroscopically determined scattering lengths.  This clearly indicates that exchange symmetry is maintained between the $|1\rangle$ and $|2\rangle$ states.  A similar measurement was made on the $|F=1,m_{f}=0\rangle \rightarrow |F=2,m_{f}=0\rangle$ transition by Fertig and Gibble \cite{Fertig2000}.

\begin{figure}
\leavevmode
\epsfxsize=3.375in
\epsffile{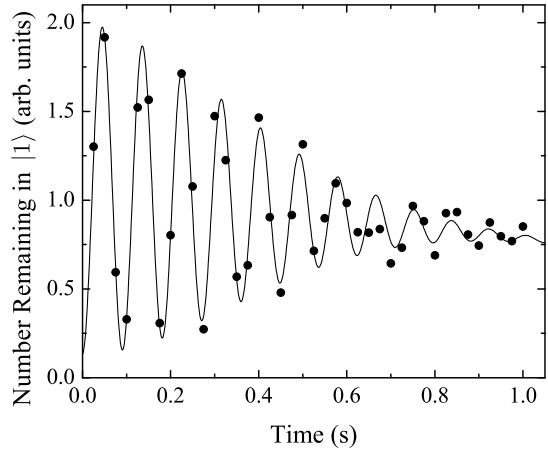}
\caption{\label{fig:chirp}  A data set of Ramsey fringes probing for frequency shifts as a function of coherence.  For this measurement normal clouds at a temperature of 480 nK and a peak density of $3.2\times10^{13}$ cm$^{-3}$ were used.}
\end{figure}

As a thought experiment, imagine distinct thermal populations of $|1\rangle$ and $|2\rangle$ atoms, separately prepared, then mixed together, with the energy of interaction (proportional to $\alpha^{nc}_{12}$) measured for instance calorimetrically.  Surely in this case the density fluctuations in state $|1\rangle$ and in state $|2\rangle$ would be uncorrelated, and $\alpha^{nc}_{12}$ would be determined to be 1, not 2.  We lack the experimental sensitivity to make such a calorimetric measurement, and our Ramsey-fringe method of measuring energy differences obviously would not work for incoherent mixtures.  We speculated, however, that if $\alpha^{nc}_{12}=2$ for coherent superpositions, and if $\alpha^{nc}_{12}=1$ for incoherent mixtures, then for partially decohered samples, $\alpha^{nc}_{12}$ would take on some intermediate value.  So by performing a measurement similar to that in Fig.~\ref{fig:fo2} we might expect to see a more negative slope for a partially decohered sample; alternatively a frequency chirp in the Ramsey fringes may be seen as the sample decoheres.

We probed the time evolution of $\alpha^{nc}_{12}$ in a way similar to Fig.~\ref{fig:fo2}; however rather than varying $f$ we set $f\simeq0.8$ then measured $\nu_{12}$ with long interrogation times, looking for a frequency chirp as the fringe contrast decreased.  This method has the advantage that there is a relatively small $|2\rangle$ state population, so effects arising from $|2\rangle$ loss are minimized.  Seven data sets were taken for this measurement; an example is shown in Fig.~\ref{fig:chirp}.  By allowing a linear frequency chirp in the fit of the Ramsey fringes, the frequency shift can be constrained to $-0.2(3)$ Hz by the time the fringe contrast has reduced to $1/e$ \cite{ChirpCorrect}.  However if we hypothesize that $\alpha^{nc}_{12}$ goes from $2$ to $1$ linearly as fringe contrast goes from $100\%$ to $0\%$ we would expect a frequency shift of $-20(2)$ Hz as the fringe decayed, while the experimental limit is 40 times smaller.  Clearly this appealing but unrigorous model is far too naive.

\begin{figure}
\leavevmode
\epsfxsize=3.375in
\epsffile{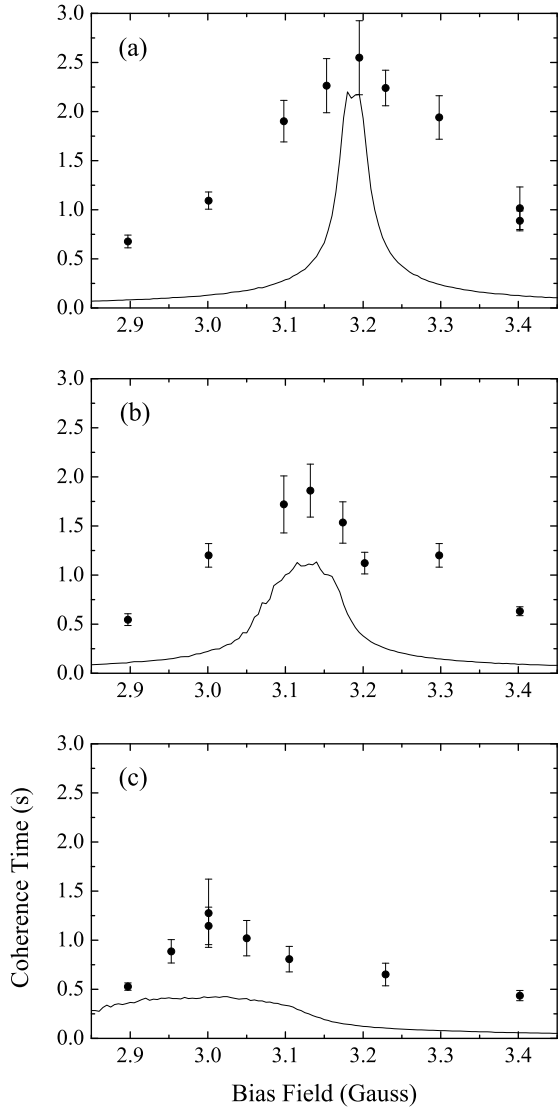} 
\caption{\label{fig:t2}  Coherence times in a normal cloud.  The temperature was $\sim480$ nK and the normal cloud peak density for each plot is: (a) $4\times10^{12}$ cm$^{-3}$; (b) $1.3\times10^{13}$ cm$^{-3}$; (c) $3.2\times10^{13}$ cm$^{-3}$.  The solid line in each plot corresponds to the $1/e$ times obtained from a numerical simulation (see text).  The noise in the simulation data is due to the finite number of particles used and the random initialization.}
\end{figure}

Ramsey spectroscopy not only allows us to probe the energy difference between the two states, but also permits the measurement of the coherence between the two states.  Coherence measurements were performed using the same time domain method used to measure $\nu_{12}$; however, interrogation times were extended until the Ramsey fringe contrast was lost.  The resulting data was fit to a $e^{{-(t/\tau)}^{2}}$ decay, where $\tau$ is the coherence time.  Additionally, fitting allowed for both a loss in total atom number and a linear frequency chirp of $\nu_{12}$.  To ensure that the $1/e$ atom loss times were much longer than the coherence times fractional transfers ($f\simeq0.8$) to the $|2\rangle$ state were used for the high density data points.  The results of coherence measurements for different magnetic fields and at three different densities are shown in Fig.~\ref{fig:t2}.

We expect that the primary source of decoherence is inhomogeneity in $\nu_{12}$ across the cloud, to which both the differential Zeeman and collisional shifts contribute.  The collisional shift scales directly with density, and thus provides a Gaussian-shaped $\nu_{12}$ profile across the cloud \cite{lewan2002}.  Inhomogeneity arising from the differential Zeeman effect depends on the bias field.  The magnitude of the magnetic field near the bottom of the trap can be written as $B(z)= \frac{B''}{2} z^2 + B_{bias}$, thus the inhomogeneity due to the Zeeman shift is $\nu_{12}\propto \frac{B''^2}{4} z^4 + (B_{bias}-B_0)B'' z^2$.  By setting $B_{bias}(>,<,=)B_0$ the curvature of $\nu_{12}$ can be adjusted to be positive, negative, or nearly zero, respectively.  This allows inhomogeneity due to the collisional shift to be roughly cancelled by an opposing Zeeman inhomogeneity \cite{lewan2002}.  This cancellation can be seen in Fig.~\ref{fig:t2}; as the density is increased from (a) to (c), the inhomogeneity induced by the collisional shift increases, so that a larger opposing Zeeman inhomogeneity is necessary for cancellation.  Therefore the bias field for peak coherence time decreases as cloud density increases.

In an attempt to compare the measured decoherence times to the known spatial inhomogeneity of the transition frequency, we performed the following numerical simulation: consistent with a Maxwell-Boltzman distribution, we randomly assign initial positions and velocities to ten thousand simulated atoms.  Ignoring the effects of collisions, we calculate the  three-dimensional trajectory of each atom for several simulated seconds, keeping track of the time integral of $\nu_{12}$ along the trajectory.  At each point in time, we calculate the spatially integrated transverse magnetization and, as inhomogeneities cause this magnetization to wash out, find the time it takes the integrated transverse magnetization to reduce to $1/e$ of its original value \cite{simT2}.  This model should correctly account for the effect of motional averaging except that all collisional  effects are explicitly excluded.  The resulting modelled damping times are plotted as a solid line along with the experimentally measured damping times in Fig.~\ref{fig:t2}.  While the model does a reasonable job predicting the value of the bias field for which the coherence time peaks, it consistently underestimates (in some cases by a factor of eight) the actual value of the coherence time.  Our model neglects both velocity-changing collisions and the exchange-type collisions that lead to spin waves; it appears that these effects contribute significantly to preserving coherence across the trapped atom cloud.

The extreme aspect ratio of our trapped cloud complicates a proper quantitative analysis of the effects of collisions on coherence.  Along the axial direction, our previous work has shown that the effects of spin-waves are to keep local magnetization across the cloud from straying too far from its spatially averaged value [see in particular Fig. 2(a) of Ref.~\cite{mcguirk2002}]. In the radial directions, the motional oscillation frequency exceeds the mean-field exchange frequency in the cloud, and thus the effects of spin-waves are probably not relevant. On the other hand, velocity-changing collisions are likely important -- in their absence, atoms with small transverse energies would stay near the axis of the trap, while atoms with large transverse energies would preferentially sample the larger magnetic fields and lower densities further from the axis.  Velocity-changing collisions which re-randomize the transverse trajectories of these different classes of atoms (before they have a time to accumulate a radian or more of relative phase-difference) will serve to extend the coherence time of the sample. In the highest density data set presented [Fig.~\ref{fig:t2}(c)] the mean elastic collision rate was 74 Hz, which should be compared the measured coherence times of around 0.5 to 1.2 seconds.

It is interesting to consider the usefulness of magnetically trapped atoms for precision metrology. Peak coherence times of approximately 2.5 s were realized with cold, low density samples.  An interrogation time of 2.5 s  provides a 0.2 Hz linewidth, which naturally leads one to consider using such a system for precision measurements.  By working at $B_{bias}=B_0$, coherence times are slightly reduced.  However, perturbations of $\nu_{12}$ due to magnetic field fluctuations become very small, on the order of 4 mHz for current typical experimental conditions.  With careful design of the confinement coils, the current supply, and magnetic shielding of external fields, it should be possible to suppress fluctuations of the bias field below 1 mG, reducing Zeeman-induced frequency shifts to below 0.1 mHz.  Perturbations of $\nu_{12}$ due to the cold collision shift are more significant; shot-to-shot density fluctuations will introduce frequency noise, so it is advantageous to work with the minimum possible density.  However as atom number is reduced the maximum signal-to-noise ratio will decrease due to shot-noise \cite{Itano1993}, therefore the optimum strategy is to work with an atom number such that the frequency uncertainty due to shot-noise is on the order of the uncertainty due to density fluctuations.

For example a normal cloud of 400 nK and a peak density $1.5\times10^{12}$ cm$^{-3}$ has $6\times10^4$ atoms.  The shot-noise-limited signal-to-noise ratio is then 245:1.  With an interrogation time of 1 second the single-shot statistical uncertainty is then 0.65 mHz; including the effects of decoherence and atom loss will increase this to approximately 0.9 mHz.  Assuming that shot-to-shot number fluctuations are also shot-noise limited, then statistical uncertainty from the density shift is 0.84 mHz.  Combining these gives a total single-shot uncertainty of 1.24 mHz; with our current 30 second evaporation time, the duty-cycle is such that an absolute precision of 6.8 mHz$/\sqrt{\mbox{Hz}}$ can be realized.  This corresponds to a relative precision of $1\times10^{-12}$ $/\sqrt{\mbox{Hz}}$, which in terms of measurement precision does not reach the level of atomic fountain clocks.  This system however has the advantage that small energy shifts can be measured in a compact, stationary spatial position; a 400 nK cloud occupies approximately only a $1040\times32\times32$ $\mu$m region of space.  It is  certainly feasible to perform spectroscopy at this level within $100$ $\mu$m of a surface, which might allow the measurement of short range atom-surface interactions.

In future work we plan on measuring coherence times in finite temperature condensates to study the role of the normal cloud in decoherence.  We anticipate this system will be quite rich due to the existence of spin waves in the normal component, phase separation in the condensate, and the interaction between the two.  The complexity of this interaction may partially account for the anomalous density shift of the condensate seen in the hydrogen experiments \cite{Hcoldc,Kleppner2002}.

We have demonstrated precise spectroscopy in an ultra-cold magnetically trapped gas.  This permitted measurement of the cold collision shift in both a condensate and a normal cloud, allowing a probe of the quantum statistics of the system.  Working at low densities minimizes the effect of the collisional shift, allowing long coherence times and precise determination of $\nu_{12}$; however the measurement of any quantity not related to atom-atom interactions will at some level be limited by this shift.  An intriguing alternative would be to instead use a fermionic atom, which should have no collisional shift, and thus density induced frequency noise will not be an issue.  On the other hand, the lack of collisions may also lead to more rapid decoherence: collisions appear to preserve our bosonic system from the decohering effects of spatial inhomogeneity.

We acknowledge useful conversations with the other members of the JILA BEC collaboration.  This work was supported by grants from the NSF and NIST.


\begin{thebibliography}{10}
\bibitem{fspinwave}
At $n=10^{13}$cm $^{-3}$ $\nu_{\mbox{spin wave}}=5$ Hz, so interrogation times of 20 ms are acceptable.
\bibitem{calibration}
The Thomas-Fermi radius (and to a lesser extent the critical temperature) are of course interaction dependent, and therefore using measurements of these quantities to calibrate densities in order to characterize interaction strengths is somewhat circular. Our directly measured value of $\alpha^{nc}_{ii}/\alpha^{c}_{ii}=2.1(2)$, on the other hand, is based predominantly on measured spatial extent of the cloud and on an absolute determination of the absorption depth, and therefore it is only weakly model dependent.  We quote the ``worst-case corrected value" of $\alpha^{nc}_{ii}/\alpha^{c}_{ii}=1.7(2)$ in order to put a reasonable limit on the size of possible systematic errors, and not because we feel it is the preferred value.
\bibitem{ExchangePulse}
When we changed the length of the first Ramsey pulse, the second pulse length was also changed such that the two pulses combined to form a $\pi$ pulse, which gives us the maximum signal-to-noise.
\bibitem{ChirpCorrect}
The quoted chirp limit of -0.2(3) Hz includes a correction to account for a small frequency change associated with decay of total atom number during the measurement time.  For each data set the number decay was found by allowing an exponential decay in the fit of the Ramsey fringes (not to be confused with the decay in Ramsey fringe contrast associated with decoherence), resulting in a correction amounting to -0.74 Hz on average.
\bibitem{simT2}
Fitting the decay of the integrated transverse magnetization to $e^{{-(t/\tau)}^{2}}$, the same functional form as we fit to the actual experimental data, gives values typically within $5\%$, and at worst $12\%$, of the $1/e$ time.
\end{thebibliography}
\end{document}